\documentstyle[jkas]{article}

\runningauthor {KANG} 
\runningtitle{ELECTRON~SPECTRUM~AT~PLANE~SHOCKS} 

\month{April} \year{2011} \volume{44} \issueno{2}
\beginpage{1}\endpage{10}
\date{Received February 12, 2011; Accepted March ??, 2011}

\def\eg{{\it e.g.,~}}
\def\ie{{\it i.e.,~}}
\def\kms{~{\rm km~s^{-1}}}
\def\cm3{~{\rm cm^{-3}}}
\def\yrs{~{\rm yrs}}
\def\muG{~{\mu\rm G}}

\begin{document}

\title{ENERGY SPECTRUM OF NONTHERMAL ELECTRONS ACCELERATED AT A PLANE SHOCK} 

\author{Hyesung Kang} 

\address{ Department of Earth Sciences, Pusan National University, Pusan  609
-735, Korea\\
 {\it E-mail : hskang@pusan.ac.kr}}

\address{\normalsize{\it (Received February 12, 2011; Accepted March ??, 2011)}}
\offprints{Hyesung Kang}
\abstract{
We calculate the energy spectra of cosmic ray (CR) protons and electrons 
at a plane shock with quasi-parallel magnetic fields,
using time-dependent, diffusive shock acceleration (DSA) simulations, 
including energy losses via synchrotron emission and Inverse Compton (IC) 
scattering. A thermal leakage injection model and a Bohm type diffusion 
coefficient are adopted. 
The electron spectrum at the shock becomes steady after the DSA energy
gains balance the synchrotron/IC losses, and it cuts off at the equilibrium
momentum $p_{\rm eq}$.
In the postshock region the cutoff momentum of the electron spectrum
decreases with the distance from the shock due to the energy losses 
and the thickness of the spatial distribution of electrons scales as $p^{-1}$. 
Thus the slope of the downstream integrated spectrum steepens by one power
of $p$ for $p_{\rm br}<p<p_{\rm eq}$, where the break momentum decrease
with the shock age as $p_{\rm br}\propto t^{-1}$.
In a CR modified shock, both the proton and electron spectrum exhibit 
a concave curvature and deviate from the canonical test-particle power-law,
and the upstream integrated electron spectrum could dominate over the
downstream integrated spectrum near the cutoff momentum. 
Thus the spectral shape near the cutoff of 
X-ray synchrotron emission could reveal a signature of nonlinear DSA.
}

\maketitle

\section{INTRODUCTION}

Diffusive shock acceleration (DSA) is widely accepted
as the primary mechanism through which cosmic rays (CRs) are
produced in a variety of astrophysical environments \citep{blaeic87, dru83}.
Detailed nonlinear treatments of DSA predict that
a small fraction of incoming thermal particles can be injected into 
the CR population, and accelerated to very high energies through their 
interactions with resonantly 
scattering Alfv\'en waves in the converging flows across collisionless shocks
\citep[e.g.][]{bkv02,bkv09,kjg02}.

Since DSA operates on relativistic particles with the same rigidity 
($R=pc/Ze$) in exactly the same manner, both electrons and protons are 
expected to be accelerated at shocks. 
Indeed, multi-band observations of nonthermal radio to $\gamma$-ray
emissions from several supernova remnants have confirmed the acceleration
of CR electrons and protons up to 100 TeV
\citep[e.g.][]{vbk05, parizot06,abdo10}.
On the other hand, postshock thermal electrons need to be pre-accelerated 
before they can be injected into Fermi process, 
since electrons have smaller gyroradius than protons by a factor of 
$(m_e/m_p)^{1/2}$  
\citep{reynolds08}.
This results in a much smaller injection rate for electrons, \ie
the CR electron-to-proton ratio is estimated to be $K_{e/p} \sim 10^{-4}-10^{-2}$,
which is not yet constrained accurately by plasma physics.
Moreover, electrons lose energy primarily by synchrotron emission and Inverse 
Compton (IC) scattering, and so their energy spectrum has a cutoff
much lower than that of the proton spectrum.

In a strong, non-relativistic shock, DSA theory predicts that the power-law
distribution function for electrons, $f_e(p)\propto p^{-q}$ with $q \approx 4$, 
which translates into the synchrotron flux, $S_{\nu} \propto \nu^{-\alpha}$
with $\alpha = (q-3)/2 \approx 0.5$.
For a unresolved, optically-thin source, we should observe the volume integrated emission
spectrum that steepens to $\nu^{-(q-2)/2}\propto \nu^{-1}$, 
because the volume integrated electron 
spectrum steepens to $F_e(p) \propto p^{-(q+1)}$ above a break momentum, $p_{\rm br}$,
due to synchrotron/IC losses. 
Such characteristics of nonthermal electrons and their synchrotron emission 
have been explored extensively in several previous studies
\citep[e.g.][]{webb84, heavens87, bkv02, za07}. 

\citet{webb84} and \citet{heavens87} considered {\it steady-state} solutions 
for the electron spectrum accelerated at a plane shock in the case of constant, 
momentum-independent diffusion coefficient, while \citet{za07} generalize it to 
the case of momentum-dependent diffusion coefficient.
In the present paper, we performed time-dependent numerical simulations in which
DSA of protons and electrons are followed along with synchrotron/IC losses
for the shock parameters relevant for typical young supernova remnants (SNRs).
A thermal leakage injection model and a Bohm-type diffusion coefficient 
($\kappa(p) \propto p$) are adopted.
Even though the CR proton spectrum extends to ever higher momentum with time,
the electron spectrum at the shock approaches to time-asymptotic states and
can be compared with the steady-state solution of \citet{za07}.
We also demonstrate how the nonlinear feedback of the CR pressure affects
the electron spectrum in a CR modified shock. 

In the next section we describe the numerical method and model parameters.
Some analytic estimates for the various features in the electron spectrum
will be  presented in Section 3. The simulation results will be discussed in Section 4, 
followed by a brief summary in Section 5.

\section{NUMERICAL METHOD}¤

\subsection{CRASH code for DSA}

Here we consider the CR acceleration at quasi-parallel shocks
where the magnetic field lines are parallel to the shock normal.
So we solve the standard gasdynamic equations with CR proton pressure terms 
added in the conservative, Eulerian formulation for one dimensional
plane-parallel geometry.
The basic gasdynamic equations and details of the CRASH (Cosmic-Ray Amr SHock) code
for one-dimensional plane-parallel geometry can be found in \citep{kjg02}. 

We solve the following diffusion-convection equations
for the pitch-angle-averaged phase space distribution function 
for CR protons, $f_p(x,p,t)$, and for CR electron, $f_e(x,p,t)$ 
\citep{skill75}:
\begin{eqnarray}
{\partial g_p\over \partial t}  + (u+u_w) {\partial g_p \over \partial x}
= {1\over{3}} {\partial \over \partial x} (u+u_w)
\left( {\partial g_p\over \partial y} -4g_p \right) \nonumber\\
+ {\partial \over \partial x} \left[\kappa(x,y)
{\partial g_p \over \partial x}\right]
\label{dcp}
\end{eqnarray}

\begin{eqnarray}
{\partial g_e\over \partial t}  + (u+u_w) {\partial g_e \over \partial x}
= {1\over{3}} {\partial \over \partial x} (u+u_w)
\left( {\partial g_e\over \partial y} -4g_e \right) \nonumber\\
+ {\partial \over \partial x} \left[\kappa(x,p)
{\partial g_e \over \partial x}\right]
+ p {\partial \over {\partial y}} \left( {b\over p^2} g \right) ,
\label{dce}
\end{eqnarray}
where $g_p= p^4 f_p$, $g_e=p^4 f_e$, $y=\ln(p)$. 
Here the particle momentum is expressed in units of $m_pc$ 
and so the spatial diffusion coefficient, $\kappa(x,p)$, has the same form 
for {\it both protons and electrons}.
The cooling term $b(p)=-dp/dt$ takes account for electron
synchrotron/IC losses. 

The velocity $u_w$ represents the effective relative motion of
scattering centers with respect to the bulk flow velocity, $u$.
The mean wave speed is set to the Alfv\'en speed, \ie
$u_w = v_A = B/ \sqrt{4\pi \rho}$ in the upstream region.
This term reflects the fact that
the scattering by Alfv\'en waves tends to isotropize
the CR distribution in the wave frame rather than the gas frame \citep{bell78}.
In the postshock region, $u_w = 0$ is assumed, since the Alfv\'enic
turbulence in that region is probably relatively balanced \citep{jon93}.
This reduces the
velocity difference between upstream and downstream scattering centers
compared to the bulk flow, leading to less efficient DSA.
Hereafter we use the subscripts '0', '1', and '2' to denote
conditions far upstream of the shock, immediately upstream of the
gas subshock and immediately downstream of the subshock, respectively.

The dynamical effects of the CR proton pressure are included
in the DSA simulations,
while the CR electrons are treated as test-particles.
In equations (\ref{dcp}) and (\ref{dce}) the spatial advection terms are solved by the 
{\it wave-propagation} algorithm, while the diffusion term
is solved by the Crank-Nicholson scheme \citep{kjg02}.

The radiative cooling term in equation (\ref{dce}) is treated by
the operator splitting method, in which the 
following advection equation in the momentum $y$ space is solved:
\begin{equation}
{\partial g_e\over \partial t}+ V \cdot  {\partial g_e\over \partial y} = 0, 
\label{gecool}
\end{equation}
where the advection speed is $V= - [b(p)/p^2]\cdot p = -D B_{\rm e}^2 p$.
Here the synchrotron cooling constant is defined as 
$D\equiv 4 e^4/(9m_e^4c^6)$ (in cgs units),
and $B_{\rm e}= (B + B_r^2)^{1/2}$ is the ``effective'' magnetic field 
strength and $B_r^2/8\pi$ is the energy density of ambient 
radiation field \citep{long94}. 
The downward advection in momentum introduces an additional Courant condition for time step, 
$\Delta t < \Delta y /|V|$.

\subsection{Model Parameters}

Here we consider a plane-parallel shock with $u_s=3000\kms$,
propagating into the upstream medium with the following properties:
gas temperature, $T_0=10^6$ K, sound speed, $c_s=150 \kms$ (sonic Mach number, $M=u_s/c_s=20$), 
and hydrogen number density, $n_0= 1 {\rm cm}^{-3}$. 
Because high-resolution X-ray observations of several young SNRs 
indicate the presence of magnetic fields as strong as a few $100 \mu$G 
downstream of the shock (\eg Parizot \etal 2006),
we adopt $B_0=30\muG$ for the upstream field strength, which is stronger
than the mean ISM field of 5-8 $\muG$.·
The upstream Alfv\'en speed is $v_A=B_0/\sqrt{4\pi\rho_0}= 65.5 \kms$
(Alfv\'enic Mach number, $M_A=u_s/v_A=45.8$) and so $v_A/c_s=0.44$.
This set of parameters may represent young supernova remnants
in the hot interstellar medium with self-amplified magnetic fields \citep{kang10}.

The physical quantities are normalized, both in the numerical code and in
the plots below, by the following constants:
\begin{eqnarray}
u_o = 3000 \kms, \nonumber\\
\kappa_o= 1.04\times 10^{27} {\rm cm^2 s^{-1}} ,\nonumber\\
r_o= \kappa_o/u_o = 3.47\times 10^{18} {\rm cm}, \nonumber\\
t_o= \kappa_o/u_o^2 = 366 {\rm years}. \nonumber
\end{eqnarray}
Note that the calculations depend on the density $\rho_0$ only through
Alfv\'en speed.  So for the cases where Alfv\'enic drift is ignored,
the density parameter $n_0$ is arbitrary. 

Since we consider only quasi-parallel shocks, the magnetic field is passive in the simulations.
However, theoretical studies have shown that efficient magnetic field 
amplification via resonant and non-resonant wave-particle interactions 
is an integral part of DSA (Lucek \& Bell 2000, Bell 2004),
and the evolution of magnetic field strength
is important in modeling the diffusion coefficient and synchrotron cooling.
For simplicity, we assume the local magnetic field strength scales
with gas density as $B(x,t)= B_0 [\rho(x,t)/\rho_0]$,
which represents enhancement of resonant Alfv\'en wave amplitudes through compression.

We adopt a Bohm-type diffusion coefficient
that includes a weaker non-relativistic momentum dependence,
\begin{equation}
\kappa(x,p) = \kappa^* \cdot \left({p \over {m_p c}} \right) \left [{\rho(x) \over \rho_0}
\right]^{-1},
\label{kappa}
\end{equation}
where the coefficient $\kappa^*= m_p c^3/(3eB_0)=1.04\times 
10^{21} {\rm cm^2 s^{-1}} (B_0/30\muG)^{-1}$. 
The density dependence approximately accounts for the compressive 
amplification of Alfv\'en waves.

Also we adopt a phenomenological injection model, 
in which particles above a certain injection momentum $p_{\rm inj}$
cross the shock and get injected to the CR population
\citep{kjg02}.
The ``effective'' injection momentum can be approximated by
\begin{equation}
p_{\rm inj} \approx 1.17 m_p u_2 \left(1+ {1.07 \over \epsilon_B} \right),
\label{pinj}
\end{equation}
where $p_{\rm th}= \sqrt{2m_p k_B T_2}$ is the thermal peak momentum of
the immediate postshock gas with temperature $T_2$
and $k_B$ is the Boltzmann constant \citep{kr10}. 
One free parameter controls this function;
$\epsilon_B = B_0/B_{\perp}$, the ratio of
the general magnetic field along the shock normal, $B_0$, to
the amplitude of the postshock MHD wave turbulence, $B_{\perp}$.
Since we are not interested in the absolute amplitude of the electron
spectrum in this study,
protons and electrons are injected in the same manner in the simulations, ignoring
the fact that postshock thermal electrons have a much smaller gyroradius.
In other words, we take effectively the electron-to-proton ratio, $K_{e/p}=1$.

\section{ANALYTIC ESTIMATES}¤

\subsection{Basics of Diffusive Shock Acceleration}

In the test-particle regime of DSA theory, the postshock CR spectrum
takes the power-law form of $f(p) \propto p^{-q}$, where the spectral index,
$q$, depends on the velocity jump across the shock:
\begin{equation}
q={{3(u_1-v_A)}\over u_1-v_A-u_2}
\label{qtp}
\end{equation}
where $u_1=u_s$ and $u_2$ are the upstream and downstream speed, respectively.
Alfv\'enic drift of self-generated waves in the upstream region reduces the velocity jump
that the particles experience across the shock, which in turn softens
the CR spectrum beyond the canonical test-particle slope.

The mean acceleration time for a particle to reach $p$ from an injection momentum
$p_{\rm inj}$
in the test-particle limit of DSA theory is given by \citep{dru83}
\begin{equation}
t_{\rm acc}(p) = {3\over {u_1-v_A-u_2}} \int_{p_{\rm inj}}^{p}
\left({\kappa_1\over {u_1-v_A}} + {\kappa_2\over u_2} \right) {dp' \over p'}.
\end{equation}
Then the mean acceleration time scale is estimated by
\begin{equation}
t_{\rm acc} (p) 
\approx F {\kappa^* \over u_s^2} \left({p \over{m_pc}}\right),
\label{tacc}
\end{equation}
where $F= 3\sigma(2-M_A^{-1})/(1-M_A^{-1})(\sigma -1 -\sigma M_A^{-1})$,
$\sigma = u_1/u_2=\rho_2/\rho_1$ is the shock compression ratio,
and $M_A=u_1/v_A$ is Alfv\'en Mach number \citep{kr10}.
In the limit of large $M_A$, 
the factor $F \approx 6 \sigma/(\sigma-1)= 2q $, and $F\approx 8$ for $\sigma=4$.
So the maximum momentum accelerated by the shock age of $t$ can be 
estimated as 
$p_{\rm max}/m_pc \approx (u_s^2/ F \kappa^*) t $,
which represents the cutoff momentum in the proton spectrum.
For typical supernova remnant shocks, it becomes
\begin{equation}
 {p_{\rm max} \over {m_pc}} \approx 3.3\times 10^{5}
\left({u_s \over {3000 \kms}}\right)^2 \left({t \over {10^3 \yrs}}\right) 
\left({B_0 \over 30\muG}\right). 
\label{pmax}
\end{equation}

\subsection{Radiative Cooling}
Relativistic electrons lose energy primarily through combined synchrotron and
IC radiation with the cooling rate $b(p)$.
So the cooling time scale is
\begin{eqnarray}
t_{\rm rad} (p) = { p \over {b(p)}} =  \left(D B_{\rm e}^2 ~p \right)^{-1} \nonumber\\
=  1.33\times 10^{6} \yrs \left({B_{\rm e} \over {100 \muG}}\right)^{-2} \left({p \over {m_pc} }\right)^{-1}.
\label{trad}
\end{eqnarray}
This can be rewritten in term of the electron Lorentz factor, 
$\gamma_e= p/m_e c$ as
\begin{equation}
t_{\rm rad}(\gamma_e)= 2.45\times 10^{2} \yrs \left({B_{\rm e} \over {100 \muG}}\right)^{-2}
\left({\gamma_e \over 10^7}\right)^{-1}.
\end{equation}

The electron cutoff energy can be estimated from the equilibrium 
condition that the momentum gain per cycle by DSA 
is equal to the synchrotron/IC loss per cycle,
\ie $\langle \Delta p \rangle_{\rm DSA}+\langle \Delta p \rangle_{\rm rad}$=0.
Following the derivation given by \citet{webb84},
\begin{equation}
\langle \Delta p \rangle_{\rm DSA}= {4p \over 3} {(u_1-u_2-v_A)\over v},
\end{equation}
\begin{equation}
\langle \Delta p \rangle_{\rm rad}= - D p^2 (B_{\rm e,1}^2 \Delta t_1 + B_{\rm e,2}^2 \Delta t_2), 
\end{equation}
where $v$ is the particle velocity,
and $B_{\rm e,1}^2= B_1^2 + B_r^2$ and $B_{\rm e,2}=B_2^2 + B_r^2$.
The upstream and downstream residence time per cycle are 
$\Delta t_1 = 4\kappa_1/(u_1-v_A) v$ and 
$\Delta t_2 = 4\kappa_2/u_2 v$, respectively.
For $\kappa_1=\sigma \kappa_2= (c^2/3 e B_1)\cdot p$,
the equilibrium momentum for electrons is given in cgs units by
\begin{equation}
p_{\rm eq}= {m_e^2 c^2 u_s \over \sqrt{4e^3q/(1-M_A^{-1})27}}
\left({B_1 \over {B_{e,1}^2/(1-M_A^{-1}) + B_{e,2}^2}}\right)^{1/2}.~~~ 
\end{equation}
If we assume $B_2=\sigma B_1$, $B_1=B_0$, $B_r=0$, $q=4$, $\sigma=4$, this becomes
\begin{equation}
{p_{\rm eq} \over {m_pc}} \approx 2.44\times 10^4 
 \left({B_0 \over 30 \muG}\right)^{-1/2} \left({u_s \over {3000 \kms}}\right), 
\label{peq}
\end{equation}
which does not depend on the shock age $t$ unlike the proton
cutoff momentum $p_{\rm max}$.

For $p \ll p_{\rm eq}$ the DSA momentum gain is much larger 
than the radiative losses, so  electrons and protons are accelerated 
in almost the same manner until the radiative loss balances the DSA gain.
This occurs at the equilibrium time, $t_{\rm eq}=t_{\rm acc}(p_{\rm eq})$, 
after which the electron spectrum at the shock becomes steady 
and cuts off at $p_{\rm eq}$.
Using equations (\ref{tacc}) and (\ref{peq}) with $F=8$, it gives
\begin{equation}
t_{\rm eq} \approx (71.7 \yrs)  
\left({B_0 \over 30 \muG}\right)^{-3/2} \left({u_s \over {3000 \kms}}\right)^{-1}. 
\label{teq}
\end{equation}

Figure 1 shows the evolution of the proton and electron spectrum {\it at
the shock position} from the test-particle DSA simulation (see next
section for details).
The electron cutoff momentum asymptotes to $p_{\rm eq}$ 
after $t\sim 0.5 t_o \approx 183 \yrs$, a few times longer 
than $t_{\rm eq}$, while the proton maximum momentum
continues to increase with time as $p_{\rm max}\propto t$. 
For $p<p_{\rm eq}$, the distribution function, $g_p(x_s,p)$ and $g_e(x_s,p)$ evolve in
exactly the same way, as expected. 

\begin{figure}[!t]
\vskip -0.5cm
\centering
\epsfxsize=8cm \epsfbox{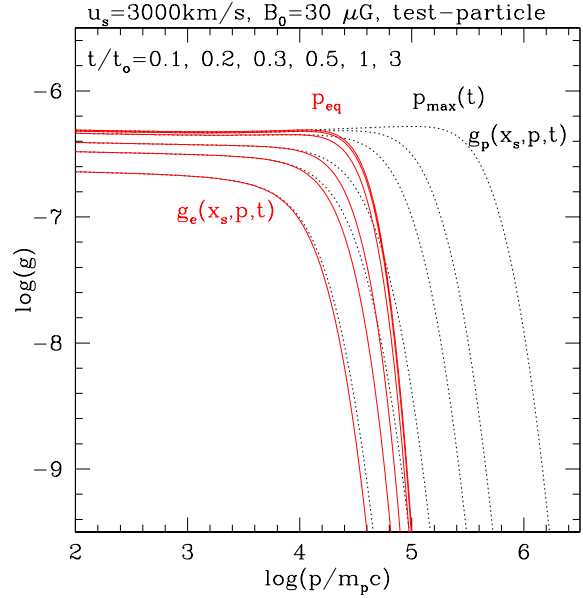}
\vskip -0.0cm
\caption{
Time Evolution of the proton and electron distribution functions, 
$g_p(x_s,p)$ (black dotted lines) and $g_e(x_s,p)$ (red solid lines),
at the shock position at $t/t_o= 0.1, 0.2, 0.3, 0.5, ~1., ~3.$ 
from the test-particle DSA simulation for $u_s=3000 \kms$, $B_0=30\muG$.
The Alf\'enic drift is not included ($v_A=0$).
}
\label{fig1}
\end{figure}

Using equation (\ref{peq}), we can estimate 
the characteristic synchrotron frequency of electrons with $p_{\rm eq}$ as
\begin{eqnarray}
\nu_c = 1.5 { {eB_2 p_{\rm eq}^2} \over {m_e^3 c^3}}
= {{81 m_e c} \over {8e^2}} {\sigma \over {q(1+\sigma^2})} u_s^2 \nonumber\\
\approx 1.1\times 10^{20} {\rm Hz} {\sigma \over {q(1+\sigma^2)}} \left({u_s \over {3000 \kms}}\right)^2,
\end{eqnarray}
which depends only on the shock speed.
So the turn-over frequency of X-ray synchrotron emission from SNRs 
provides a measure of the shock speed, but not the magnetic field strength directly.

\citet{za07} suggested that the electron spectrum at the shock position 
for a steady-state plane shock can be approximated by
\begin{equation}
f_e(x_s, p) \propto p^{-4}[1+0.523(p/p_0)^{9/4}]^2 \exp(-p^2/p_0^2),
\label{fza}
\end{equation}
in the case of Bohm diffusion with $B_2=\sqrt{11} B_0$ and $\sigma=4$.
Their cutoff momentum $p_0$ is given by
\begin{equation}
{p_0 \over {m_p c}} \approx { 1.56\times 10^5 \over {q(1+\sigma^{-1/2})} } \left({u_s \over {3000 \kms}}\right)
\left({B_2 \over {100 \muG}}\right)^{-1/2}, 
\label{p0}
\end{equation}
which is similar to, but not the same as, $p_{\rm eq}$ in equation (\ref{peq}), 

\begin{figure*}[t]
\vskip -0.5cm
\centering
\epsfxsize=13cm \epsfbox{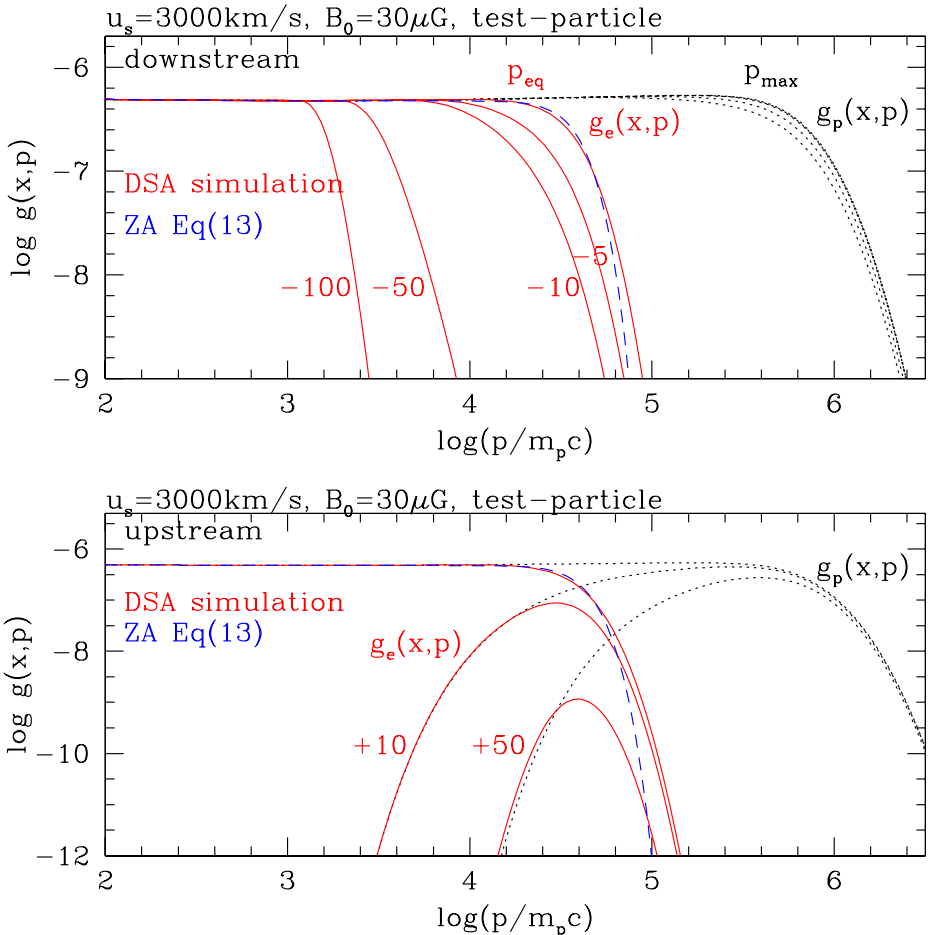} 
\vskip -0.5cm
\caption{ 
Particle spectra from the test-particle DSA simulation for 
$u_s=3000 \kms$, $B_0=30\muG$. 
{\it Top:}
Proton spectrum $g_p(x,p)$ (dotted lines) and electron spectrum 
$g_e(x,p)$ (solid lines) at 5 different locations downstream of the
shock at $t/t_o=5$ are shown.
The curves for the electron spectrum are labeled with the distance
downstream from the shock in the number of grid zones (-100, -50, -10, -5, 0). 
Note that the cooling length for electrons with $p_{\rm eq}$, $l_{\rm eq}$,
corresponds to 6.15 grid zones in the simulation.
The dashed line shows the analytic form for the electron spectrum
at the shock (Eq. [\ref{fza}]) suggested by \citet{za07}.
{\it Bottom:}
Same as above except the spectra at two locations (+10 and +50
grid zones) upstream of the shock are shown along with the spectra at 
the shock position.
}
\label{fig2}
\end{figure*}


At the shock the electron spectrum cuts off at $p_{\rm eq}$,
but in the postshock region electrons lose energy 
while being advected downstream,
and so the electron cutoff momentum moves downward in the momentum space.
The cutoff momentum, $p_{\rm cut}$, 
at the distance $d$ downstream from the shock is given by   
\begin{equation}
p_{\rm cut}(d) = {p_{\rm eq}\over {1+ d/l_{\rm eq}}},
\label{pcut}
\end{equation}
where $l_{\rm eq}=  u_2 \cdot t_{\rm rad}(p_{\rm eq}) = u_2(D B_{e,2}^2 p_{\rm eq})^{-1}$.
In the limit of $d \gg l_{\rm eq}$, 
the cutoff momentum of the electron spectrum decreases 
away from the shock as 
$p_{\rm cut}= p_{\rm eq}\cdot (l_{\rm eq}/d) \propto d^{-1}$.
This implies that the thickness of the spatial distribution
of electrons with momentum $p$ decrease as 
$\Delta x(p) = u_2 t_{\rm rad}(p) \propto p^{-1}$. 

\begin{figure*}[t]
\centering
\vskip -1.5cm
\epsfxsize=14cm \epsfbox{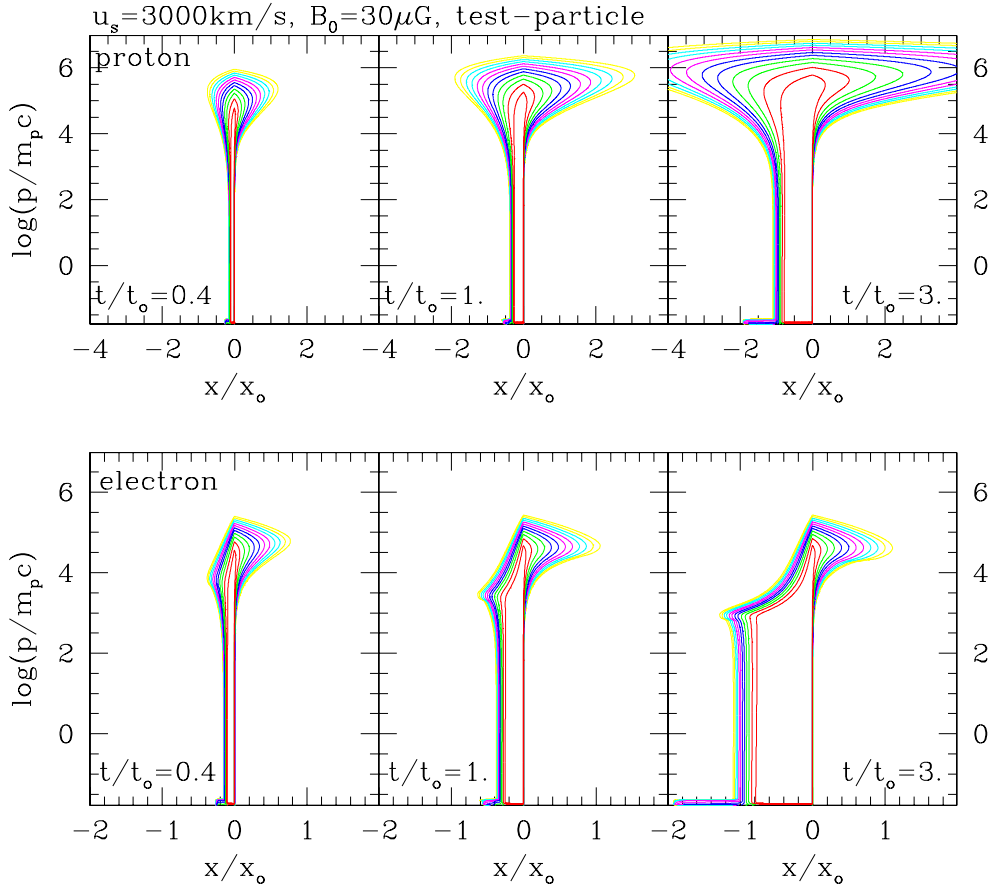} 
\vskip -1.0cm
\caption{
Proton and electron distribution function, $g_p(x,p)$ and $g_e(x,p)$,
in the phase-space at $t/t_o= 0.4, ~1, ~3$ 
for the test-particle case shown in Fig. 2.
At each contour level, the value of $g$ increases by a factor of 10. 
Note that the spatial span of the proton distribution shown here is [-4,+4],
while that of the electron distribution is [-2,+2].
}
\label{fig3}
\end{figure*}

At the shock age $t \gg t_{\rm eq}$,
the electron spectrum at the furthest downstream point
($d_{\rm adv}=u_2\cdot t$) has the cutoff at 
$p_{\rm cut} = p_{\rm br}(t) \approx p_{\rm eq} (l_{\rm eq}/u_2 t) \approx (DB_{\rm e,2}^2t)^{-1}$.
This gives 
\begin{equation}
{p_{\rm br}(t) \over {m_pc}} \approx 1.34 \times 10^3 \left({t \over 10^3 \yrs}\right)^{-1} \left({B_{\rm e,2} \over
{100 \muG}}\right)^{-2},
\label{pbr}
\end{equation}
which depends only on the postshock magnetic field strength and
the shock age, but not on the shock speed.
If the electron distribution function at the shock has the form,
$f_e(x_s,p)\propto p^{-q}$ for $p<p_{\rm eq}$,
then the downstream integrated electron spectrum is given by   
\begin{equation}
F_{e,2}(t,p)=\int_{-\infty}^{0} f_e(x,p,t) dx \propto p^{-(q+1)}, 
\end{equation}
for $p_{\rm br}<p<p_{\rm eq}$,
which was also shown in previous studies \citep[e.g.][]{heavens87, za07}.

\section{DSA SIMULATION RESULTS}

\subsection{Test-Particle Case}

\begin{figure*}[t]
\centering
\vskip -1.5cm
\epsfxsize=14cm \epsfbox{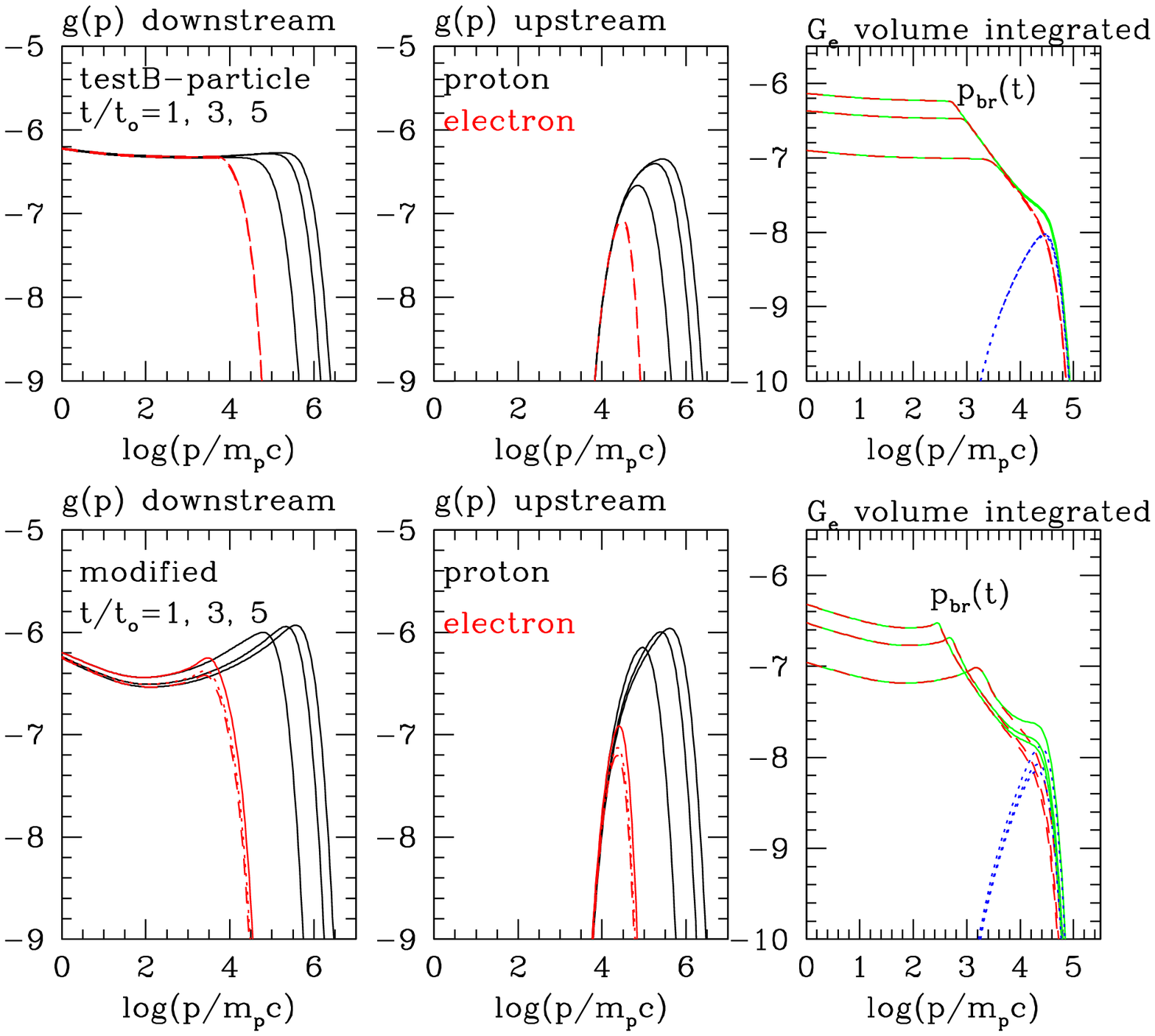} 
\vskip -1.0cm
\caption{
Top: DSA simulation results from the test-particle case: $u_s= 3000 \kms$, $B_0=30\muG$, $v_A=0$, $\epsilon_B=0.2$.
Bottom: DSA simulation results from the CR modified case: $u_s= 3000 \kms$, $B_0=30\muG$, $v_A=65.5\kms$,
$\epsilon_B=0.25$.
Proton and electron distribution functions, $g_p(x,p)$ (black solid lines) and $g_e(x,p)$ 
(red lines),
at the shock location (left panels) and at a upstream location (middle panels)
at $t/t_o=1,$ 3, 5.
The right panels show the downstream integrated electron spectrum, 
$G_{e,2}=\int_{-\infty}^{0} f_e(p)p^4 dx$ (red dashed lines), the 
upstream integrated electron spectrum, $G_{e,1}=\int_{0}^{+\infty} 
f_e(p)p^4 dx$ (blue dotted lines) and the sum of the two curves (green solid lines) at $t/t_o=1,$ 3, 5.
}
\label{fig4}
\end{figure*}

In order to make a direct comparison between the analytic estimates and
the DSA simulation results,
we first consider a test-particle case, in which the dynamical
feedback of the CR pressure is ignored. So the shock structure is
determined by the canonical shock jump conditions.
The injection parameter, $\epsilon_B=0.2$ is adopted
and the Alfv\'enic drift is ignored (\ie $v_A=0$) for this case. 

Figure 2 shows the proton and electron spectra at different locations
at $t=5t_o=1.83\times 10^3 \yrs$.
It shows that the cutoff momentum for the proton spectrum is 
$p_{\rm max}/m_p c \approx 6\times 10^5$ (Eq. [\ref{pmax}]),
while that for the electron spectrum at the shock is
$p_{\rm eq}/m_p c \approx 2.4\times 10^4$ (Eq. [\ref{peq}]).
For comparison, we also show the analytic form in equation (\ref{fza}) suggested by
\citet{za07}, which agrees well with the simulated electron spectrum at the shock,
especially near the exponential cutoff. 

The DSA operates efficiently 
only within a diffusion length, $\sim \kappa(p)/u_s$, from the shock.
Far behind the shock radiative cooling of electrons dominates over DSA, 
so the function $g_e(x,p)$ shifts downward in $\ln p$ space 
in a manner approximately described by the advection equation in (\ref{gecool}).
Because the advection speed depends on electron momentum as $V \propto p$,
the shape of the cutoff steepens as we move downstream from the shock.
So the electron spectrum well behind shock cannot be approximated by
a simple analytic form such as a power-law with a Gaussian exponential cutoff.
\citet{heavens87} also observed the same trend in the electron spectrum
behind a steady-state plane shock in the case of constant magnetic
field strength and momentum-independent diffusion coefficient (\ie
$B_1=B_2$ and $\kappa_1=\kappa_2$). 
According to equation (\ref{pcut})
the cutoff momentum is estimated to be $p_{\rm cut}/m_p c \approx
1.3\times 10^3,~ 2.5\times10^3,~ 8.8\times 10^3, ~1.3\times 10^4$ 
at 100, 50, 10, 5 grid zones, respectively, downstream from the shock.
These estimates match well the simulated electron spectra, as can be
seen in Figure 2.

In upstream region, only the highest energy particles diffuse away from the shock,
while low energy particles are swept downstream.
So there the proton spectrum $g_p(p)$ has a peak at $p_{\rm max}$,
while the electron spectrum $g_e(p)$ has a peak at $p_{\rm eq}$,
as can be seen in the bottom panel of Figure 2. 
The upstream electron spectrum has the same exponential cutoff at $p_{\rm eq}$
as the spectrum at the shock.  Again the electron spectrum has the same
shape as the proton spectrum for $p<p_{\rm eq}$.

Figure 3 shows the evolution of the phase-space distribution of $g_p(x,p)$
and $g_e(x,p)$ for the test-particle case.
At low energy, $p/m_p c < 10^3$, electron cooling is not important and
so protons and electrons have the same distribution. 
At higher energy, electrons have much narrower distribution both in $x$ and
$\ln p$ space due to radiative cooling.
It also demonstrates that the spatial thickness of the electron distribution
narrows down at higher momentum as $\Delta x \propto p^{-1}$ in the postshock 
region,
and that the cutoff momentum $p_{\rm cut}(x)$ decreases as the downstream
distance increases away from the shock. 
Although the postshock electron distribution extends downstream 
as the flow advects, the upstream electron distribution does not change
after $t/t_o\ga 1$ due to the balance between DSA and cooling.
On the contrary, high energy protons diffuse much further both downstream
and upstream as $p_{\rm max}\propto t$ increases with time.

The top panels of Figure 4 show the evolution of $g_p(p)$ and $g_e(p)$
at the shock (right panel) and at a upstream position (middle panel)
at $t= 3.7\times 10^2,~ 1.1\times 10^3, ~1.8\times 10^3 \yrs$.
Because $t\gg t_{\rm eq}=71.7 \yrs$, so the electron spectrum
has already reached the steady-state, as can be seen in the figure.
In the right panel we also plot the downstream integrated electron spectrum, 
$G_{e,2}=\int_{-\infty}^{0} g_e(p) dx$, the 
upstream integrated electron spectrum, $G_{e,1}=\int_{0}^{+\infty} g_e(p) dx$.
As discussed right after equation (16), $G_{e,2}$ (dashed lines)
 becomes a broken power-law which
steepens from $p^{-q}$ to $p^{-(q+1)}$ above $p> p_{\rm br}(t)$ 
with an exponential cutoff at the higher momentum $p_{\rm eq}\approx 2.4\times 10^4$.
We also see that the brake momentum decrease with time as
$p_{\rm br} \propto t^{-1}$.
Note that the upstream integrated spectrum $G_{e,1}$ (dotted lines) 
has reached the steady-state, 
while, for $p<p_{\rm br}$, the amplitude of the downstream spectrum $G_{e,2}$ 
increase linearly with time as the flow advects downstream.
Such a trend was seen in Figure 3 as well.
Total volume integrated spectrum (solid lines) shows a small bump near 
the cutoff momentum due to the upstream contribution,
which in turn will determine the exact shape of the cutoff of X-ray synchrotron emission.

\begin{figure*}[t]
\centering
\vskip -1.5cm
\epsfxsize=14cm \epsfbox{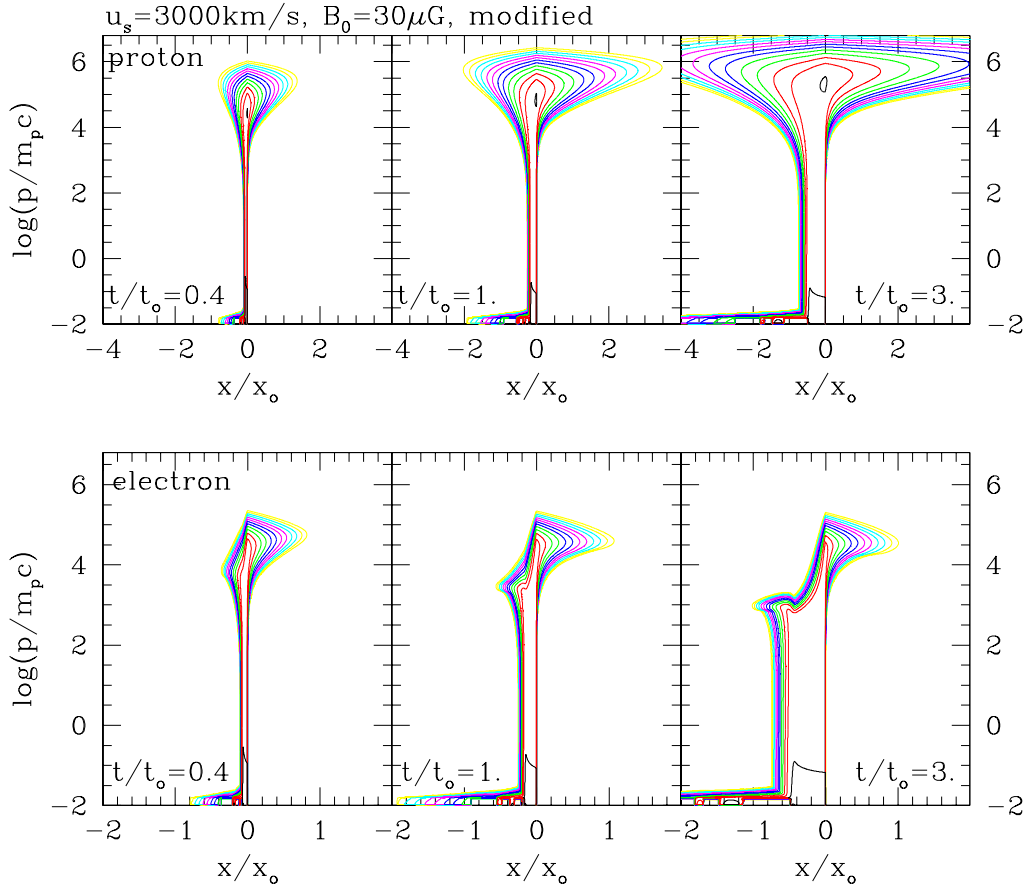} 
\vskip -1.0cm
\caption{
Proton and electron distribution functions, $g_p(x,p)$ and $g_e(x,p)$,
in the phase-space at $t/t_o= 0.4, ~1, ~3$ 
for the CR modified case shown in the bottom panels of Fig. 4.
At each contour level, the value of $g$ increases by a factor of 10. 
Note that the spatial span of the proton distribution shown here is [-4,+4],
while that of the electron distribution is [-2,+2].
}
\label{fig5}
\end{figure*}

\subsection{CR Modified Case}

With the injection parameter $\epsilon_B=0.25$, 
the CR injection and acceleration is efficient enough to modify significantly 
the shock structure.
The CR injection fraction becomes $\xi \approx 5\times 10^{-4}$, 
and the postshock CR pressure is $P_{cr,2}/\rho_0 u_s^2\approx 0.29$.
So the postshock gas density, $\rho_2/\rho_0 \approx 5.6$, is larger
than $\sigma=4$ for the gasdynamic shock.
In such CR modified shocks, the pressure from CRs diffusing upstream compresses
and decelerates the gas smoothly before it enters the subshock,
creating a shock precursor \citep{kj07}.

With the assumed momentum-dependent diffusion, $\kappa(p)$,
the particles of different momenta, $p$, experience different
compressions, depending on their diffusion length, $l_d(p) =\kappa(p)/u_s$.
The particles just above $p_{\rm inj}$
sample mostly the compression across the subshock ($\sigma_s=\rho_1/\rho_0$),
while those near $p_{\rm max}$
experience the total compression across the entire shock structure
($\sigma_t=\rho_2/\rho_0$).
This leads to the particle distribution function that
behaves as $f(p)\propto p^{-3\sigma_s/(\sigma_s-1)}$ for $p\sim p_{\rm inj}$,
but flattens gradually to $ f(p)\propto p^{-3\sigma_t/(\sigma_t-1)}$ toward
$p\sim p_{\rm max}$ \citep{krj09}.

The bottom panels of Figure 4 show the evolution of $g_p(p)$,
$g_e(p)$ at the shock position (left panel) and at a upstream location (middle
panel) and the volume integrated $G_e(p)$ (right panel) as discussed before.
The Alfv\'enic drift with $v_A=0.44 c_s ~(M_A=45)$ are considered.
Because of the development of a smooth precursor and the weaker subshock,
both $g_p$ and $g_e$ at the shock are softer than the test-particle
power-law spectra at lower momenta $p/(m_p c)<10^2$,
while they are harder at higher momenta with a cutoff at
$p_{\rm max}$ (proton spectrum) or $p_{\rm eq}$ (electron spectrum).
Thus the CR spectrum exhibits the well-known concave curvature 
between the lowest and the highest momenta.
Such concavity is reflected in the volume integrated spectrum as well,
so $G_e$ is no longer a simple broken power-law as in the test-particle case.

With the greater velocity jump ($\sigma_t=5.6$), the acceleration is more efficient
and so there are more highest energy particles in the upstream region,
compared to the test-particle case. 
As a result, the upstream integrated spectrum, $G_{e,1}$ (dotted lines)
has a more pronounced peak at $p_{\rm eq}$, compared to the test-particle case.
This introduces an additional curvature in the total $G_e$ spectrum.
In fact, $G_{e,1}$ dominates over $G_{e,2}$ near $p_{\rm eq}$, so 
the upstream contribution should determine the spectral shape 
of X-ray synchrotron emission. 
Thus the spectral slope in radio and the detail shape in X-ray
of the observed synchrotron flux can provide a measure of 
nonlinear DSA feedback.
  
Finally, Figure 5 shows the phase-space distribution of $g_p(x,p)$
and $g_e(x,p)$ for the CR modified model with high injection rate.
Because the postshock magnetic field strength, $B_2 = \sigma_t B_0=5.6B_0$,
is stronger, electrons cool down to lower energies, compared to
the test-particle case shown in Figure 3.
Enhanced cooling also reduces the thickness of the electron spatial
distribution downstream of the shock.
Again, we can see that at the highest energies of $p/m_pc > 10^4$,
the upstream electron components is more important than the
downstream component.

\section{SUMMARY}

Using the kinetic simulations of diffusive shock acceleration at a plane shock,
we calculate the time-dependent evolution of the CR proton and electron spectra,
including electronic synchrotron/IC energy losses.
Both protons and electrons are injected at the shock via thermal leakage injection
and accelerated by DSA, while electrons are treated as test-particles.
We adopt a momentum-dependent, Bohm-type diffusion coefficient 
and assume that the magnetic field strength scales with the gas density.

The proton spectrum at the shock, $g_p(x_s,p)$, and the volume-integrated proton
spectrum, $G_p(p)$ extends to $p_{\rm max}$ in equation (\ref{pmax}), which increases linearly with time. 
On the other hand, the electron spectrum at the shock, $g_e(x_s,p)$, approaches
to the time-asymptotic spectrum for the shock age $t>t_{\rm eq}$ in equation (\ref{teq}) .
In that regime, our time-dependent results with a Bohm-type diffusion are 
qualitatively consistent with the analytic solutions for a {\it stead-state} plane shock,
which were previously presented by several authors such as \citet{heavens87}
with {\it momentum-independent} $\kappa$ and \citet{za07} with {\it momentum-dependent} $\kappa(p)$.
So we will re-iterate some of the major findings discussed by those authors 
and add new insights obtained from our nonlinear DSA simulations.

1) First of all, we re-derive two characteristic momenta:  
the cutoff momentum, $p_{\rm eq}$ in equation (\ref{peq}) (for the Bohm-type
diffusion coefficient) and the break momentum, $p_{\rm br}$ in equation (\ref{pbr}).
Note that $p_{\rm eq}$ is a time-asymptotic quantity that is
achieved when the DSA energy gain balances the synchrotron/IC energy losses,
while $p_{\rm br}$ is a time-dependent quantity
that is determined by 'aging' of electrons due to synchrotron/IC cooling
downstream of the shock.

2) The time-asymptotic electron distribution function at the shock, $f_e(x_s,p)$,
has a Gaussian cutoff as $\exp(-p^2/p_{\rm eq}^2)$, 
which agrees well with the analytic form suggested by \citet{za07}. 

3) Behind the shock synchrotron/IC cooling dominates over DSA, so the electron
spectrum, $f_e(x,p)$, cuts off at progressively lower 
$p_{\rm cut}(d) \approx (u_2/D B_{e,2}^2) d^{-1} $,
which decrease with the distance, $d=x_s-x$, from the shock
and is smaller than $p_{\rm eq}$. 
This cutoff momentum is determined by the cooling rate $D B_{e,2}^2$,
independent of DSA.

3) The electron cooling can be represented by the advection of the distribution function 
$g_e(p)=f_e(p)p^4$ in $y=\ln(p)$ space with the advection speed, $V=-DB_e^2p$ (see Eq. [\ref{gecool}]).  
This causes the electron spectrum, $f_e(x,p)$,
cuts off more sharply as the distance downstream from the shock increases.

4) Because the synchrotron/IC cooling time decreases with momentum as $t_{\rm rad} \propto p^{-1}$, 
thickness of the electron distribution is inversely proportional to the momentum, \ie
$\Delta x(p) =u_2 \cdot t_{\rm rad} \propto p^{-1}$. Then the electron spectrum integrated over 
to the downstream region steepens as $F_{e,2}(p)\propto p^{-(q+1)}$
for $p_{\rm br}(t)<p<p_{\rm eq}$, when
the spectrum at the shock is $f_e(x_s,p) \propto p^{-q}$.
The break momentum decreases with the shock age as $p_{\rm br} \propto t^{-1}$ (see Eq. [\ref{pbr}]). 

5) Only highest energy electrons diffuse upstream to the distance of 
$d \sim \kappa(p_{\rm eq})/u_s$, so the upstream integrated spectrum has a much
harder spectrum than the downstream integrated spectrum and it 
peaks at $p_{\rm eq}$.

6) For a CR modified shock, both proton and electron spectra exhibit the well-known
concave curvatures.  Thus the volume integrated spectrum, $F_e(p)$, cannot be represented
by the canonical broken power-law spectrum.
In this regime, the radio synchrotron index, $\alpha$, could be steeper than 0.5
even for a high sonic Mach number. Also in the case of small Alfv\'enic Mach
number (\ie large $B_0$ and small $\rho_0$), the spectral slope could be even
steeper due to the Alfv\'enic drift effect. 
Moreover, detail analysis of the X-ray synchrotron emission near the cutoff 
frequency may provide some information about the effect of nonlinear DSA at shocks.  
Spectral characteristics of the synchrotron emission from a CR modified shock will be
presented elsewhere.
 
\acknowledgments{
This research was supported by Basic Science Research Program through 
the National Research Foundation of Korea (NRF) funded by the Ministry 
of Education, Science and Technology (2010-0016425).
}


\end{document}